# THE PÖSCHL-TELLER LIKE DESCRIPTION OF QUANTUM -MECHANICAL CARNOT ENGINE.

**E.O. Oladimeji[1a,b], S.O. Owolabi[c], J.T. Adeleke[d]**

a. Department of Physics, Federal University Lokoja (FULOKOJA), Lokoja, Nigeria.
b. Institute of Physical Research and Technology, Peoples' Friendship University of Russia (RUDN), Moscow, Russia.
c. Department of Physical Science, Joseph Ayo Babalola University (JABU), Ikeji-Arakeji, Nigeria.
d. Physics Department, Osun State University (UNIOSUN), Oshogbo, Nigeria.

## ABSTRACT

In this work, an example of a cyclic engine based on a quantum-mechanical properties of the strongly non-linear quantum oscillator described by the Pöschl-Teller [PT] model is examined. Using the [PT] model as shown in our earlier works [1–4], a quantum-mechanical analog of Carnot cycle (i.e. Quantum heat engine) has been constructed through the changes of both, the width $L$ of the well and its quantum state. This quantum heat engine has a cycle consisting of adiabatic and isothermal quantum processes. The efficiency of the quantum engine based on the Pöschl-Teller-like potential is derived and it's analogous to classical thermodynamic engines.

*Keywords: Quantum Thermodynamics, Quantum Mechanics, Carnot cycle, Quantum Heat Engine, Nano-Engine, Pöschl-Teller model.*

## 1. INTRODUCTION

With the rapid development of quantum thermodynamics, research on quantum heat engines has become a subject of renewed interest after being introduced in 1959 by *Scovil and Schultz-Dubois* [5]. Of recent, diverse efforts have been made to understand the working mechanism of Quantum heat engines (QHE) [6–12] which has led to the introduction of several Quantum analogues [13–16]. Several working substances such as the spin systems [17,18], two-level or multilevel systems [16,19], particle in a box [19,20], cavity quantum electrodynamics systems[21,22], coupled two-level systems [23], Harmonic oscillators [8,17,24–26], Pöschl-Teller Oscillator[1,4,27], etc. have been investigated by researchers.

These Quantum heat engines (QHE) are microscopic engines that are described by dynamical equations of motion that obey the laws of quantum mechanics [28]. This field considers analogies between quantum systems and macroscopic engines, with an example in the proposed model by *Bender et al* [19] of a cyclic engine based on a single quantum mechanical particle of mass $m$ confined in an infinite one-dimensional potential well of width $L$ (free particle [FP] in the box). This model replaces the role of piston in a cylinder and temperature in classical thermodynamics of the walls of the confining potential and energy as given by the pure-state expectation value of the Hamiltonian in Quantum thermodynamics respectively. This concept was useful in our earlier work where we formulated and applied the Pöschl-Teller [PT] potential in Joule-Brayton and Otto cycle [1–4], rather than the [FP] as used by *Bender et al* [19].

In this paper, we use the [PT] model to construct an adiabatic and isothermal quantum analogous process to analyze the efficiency of an idealized reversible heat engine i.e. Carnot engine. This model uses the walls of the containing potential to play the role of the piston in a cylinder containing an ideal gas and then constructs quantum mechanical equivalents of isothermal and adiabatic processes. The pressure $P$ exerted on this wall as defined by Hellman and Feynman is [1,4,29,30]:

---

[1] *Corresponding author: E. O. Oladimeji. e-mail: nockjnr@gmail.com;enock.oladimeji@fulokoja@edu.ng*

$$\hat{P}_n(L) = -\frac{\partial \hat{E}_n}{\partial L} \qquad (1)$$

where $E$ is the energy operator such that the formal relation between pressure $P$ and energy $E$ operator or Hamiltonian is $\hat{P}(\hat{x}, \hat{p}, L) = -(\partial/\partial L)H(\hat{x}, \hat{p}, L)$. The average energy replaces the role of temperature which is derived from the expectation value of the Hamiltonian.

In 1933, *G. Pöschl* and *E. Teller* introduced a family of anharmonic [PT]-potential $V(x; L) = -V(x; L) = V_0 tg^2[\alpha(L)x]$; $\alpha(L) = \pi/L$ where at $x = \pm L$ the potential becomes singular, which physically means the presence of a pair of impenetrable walls [31]. This potential allows the exact solution of one-dimensional Schrödinger equation with fully discrete positive energy levels in the coordinate $E_n(L) > 0$ [27].

$$E_n^{PT}(L) = W(L)[n^2 + \lambda(L)(2n + 1)] \qquad (2)$$

where; $W(L) = \pi^2 \hbar^2/2mL^2$ and $\lambda(L) = [(2/(\pi\zeta(L))^2 + 1]^{1/2} - 1$ as defined in [1]. The [PT] potential lies between the infinite-box [FP] and the harmonic oscillator [HO] potentials. The parameter $\lambda(L)$ thus indicates the position of the [PT] potential between the [FP] and [HO] limits[27,32] while $\zeta(L)$ is like the particle's momentum $p$ [1]. In the limit $\lambda \cong 0$, the potential becomes the infinite square well [FP] while in the opposite limit, $\lambda \to 1$ the harmonic oscillator [HO] potential emerges[32].

Since the pressure operator $P_n(L) = (s/L)E_n(L)$, where $s = 2$ therefore;

$$P_n^{PT}(L) = \frac{2W(L)}{L}[n^2 + 2\lambda(L)(n + \tfrac{1}{2})\{1 - \mu(L)\}] \qquad (3)$$

where $\mu(L) = 1 - [\lambda(L) - 1][2\lambda(L) - 1]^{-1}$.

## 2. THE CARNOT CYCLE

The classical Carnot cycle is composed of two isothermal and two adiabatic processes *(see Fig.1)* each of which is reversible.

Carnot cycle is the most efficient heat engine allowed by physical laws. As proposed by Carnot, it is an ideal mathematical model of a heat engine with an efficiency of almost 100%, it is cyclic and reversible although it is practically impossible in real engines.

To achieve a quantum description of the Carnot cycle, *Bender et al.* formulated a two-state model of a single particle confined in a one-dimensional infinite potential well and have devised a reversible cycle by changing the potential width and the quantum state which shows the possibility to construct a quantum-mechanical analog of the Carnot engine [19]. In this description, the cylinder in classical Carnot cycle is replaced by a potential well in quantum Carnot cycle likewise, the fluid (an ideal gas) and temperature $T$ in classical thermodynamics are replaced with a single quantum particle and expectation value of the Hamiltonian $E$.

Classically, the temperature $T$ and internal energy of the gas in the cylinder remains constant during an isothermal process even as the piston compresses or expands the gas, work is done yet the system remains equilibrium all through. While Quantumlly, given that the system at the initial state $\psi(x)$ of volume $L$ is a linear combination of eigenstates $\phi_n(x)$, the expectation value of the Hamiltonian remains constant as the walls of the well moves. The expansion coefficient $a_n$ changes such that $E(L)$ remain fixed as $L$ changes:

$$E(L) = \sum_{n=1}^{\infty} |a_n|^2 E_n \qquad (4)$$

where $E_n$ is the [PT] energy spectrum (1) and the coefficients $|a_n|^2$ are constrained by the normalization condition $\sum_{n=1}^{\infty} |a_n|^2 = 1$.

**Process 1: Isothermal Expansion**

Given that the piston expands isothermally such that the system is excited from its initial state $n = 1$ at point 1 (i.e. from $L = L_1$ to $L = L_2$) and into the second state $n = 2$, keeping the expectation value of the Hamiltonian constant. Thus, the state of the system is a linear combination of its two energy eigenstates:

$$\Psi_n = a_1(L)\phi_1(x) + a_2(L)\phi_2(x),$$

where $\phi_1$ and $\phi_2$ are the wavefunctions of the first and second states, respectively.

$$\Psi_n = a_1(L)\sqrt{2/L} \sin(\alpha x) + a_2(L)\sqrt{2/L} \sin(2\alpha x)$$

$$E(L) = \sum_{n=1}^{\infty} (|a_1|^2 + |a_2|^2) E_n = |a_1|^2 E_1 + |a_2|^2 E_2$$

The coefficients satisfy the condition $|a_1|^2 + |a_2|^2 = 1$. The expectation value of the Hamiltonian in this state as a function of $L$ is calculated as $E = \langle \psi|H|\psi \rangle$:

$$E(L) = W(L)\big[4 + 5\lambda(L) - (3 + 2\lambda(L))|a_1|^2\big], \qquad (5)$$

Setting the expectation value to be equal to $E_H$ i.e. $n = 1$

$$\frac{\pi^2 \hbar^2}{2mL_1^2}[1 + 3\lambda(L)] = \frac{\pi^2 \hbar^2}{2mL^2}\big[4 + 5\lambda(L) - (3 + 2\lambda(L))|a_1|^2\big]$$

$$L^2 = \frac{L_1^2}{[1+3\lambda(L)]}\big[4 + 5\lambda(L) - (3 + 2\lambda(L))|a_1|^2\big] \qquad (6)$$

The max value of $L$ is when $L = L_2$ and this is achieved in the isothermal expansion when $|a_1|^2 = 0$. Therefore, from equ. (6)

$$L_2 = L_1[4 + 5\lambda(L)/1 + 3\lambda(L)]^{1/2} \qquad (7)$$

The pressure during the isothermal expansion is:

$$P(L) = \sum_{n=1}^{\infty}(|a_1|^2 + |a_2|^2)P_n = |a_1|^2 P_1 + |a_2|^2 P_2$$

$$P_1(L) = (2W(L)/L)[4 + 5\lambda(L)\{1 - \mu(L)\} - (3 + 2\lambda(L)\{1 - \mu(L)\})|a_1|^2]$$

Considering the value $\mu(L)$ to be negligible;

$$P_1(L) = (2W(L)/L)[4 + 5\lambda(L) - (3 + 2\lambda(L))|a_1|^2] \tag{8}$$

Substituting (7) into (8);

$$[4 + 5\lambda(L) - (3 + 2\lambda(L))|a_1|^2] = \frac{L^2}{L_1^2}[1 + 3\lambda]$$

Therefore;

$$P_1(L) = \frac{\pi^2 \hbar^2}{mLL_1^2}[1 + 3\lambda] \tag{9}$$

The product $LP_1(L) = constant$. This is an exact quantum analogue of a classical *equation of state*.

**Process 2: Adiabatic Expansion**

Next, the system expands adiabatically from $L = L_2$ until $L = L_3$. During this expansion, the system remains in the second state $n = 2$ as no external energy comes into the system and the change in the internal energy equals to the work performed by the walls of the well. The expectation value of the Hamiltonian is:

$$E_L = \frac{\pi^2 \hbar^2}{2mL_3^2}[4 + 5\lambda(L)]$$

$$E_C = 2W(L)[4 + 5\lambda(L)] \tag{10}$$

The pressure $P$ as a function of $L$ is:

$$P_2 = \frac{2W(L)}{L}[4 + 5\lambda(L)\{1 - \mu(L)\}] \tag{11}$$

The product $L^3 P_2(L)$ in (11) is a constant and it is considered as the quantum analogue of the classical *adiabatic process*.

**Process 3: Isothermal Compression**

The system is in the second state $n = 2$ at point 3 and it compresses Isothermally to the initial (ground) state $n = 1$ (i.e. from $L = L_3$ until $L = L_4$) as the expectation value of the Hamiltonian remains constant. Thus, the state of the system is a linear combination of its two energy eigenstates.

$$\Psi_n = b_1(L)\phi_1(x) + b_2(L)\phi_2(x)$$

where $\phi_1$ and $\phi_2$ are the wave functions of the first and second states respectively

$$\Psi_n = b_1(L)\sqrt{2/L}\sin(\alpha x) + b_2(L)\sqrt{2/L}\sin(2\alpha x)$$

$$E(L) = \sum_{n=1}^{\infty}(|b_1|^2 + |b_2|^2)E_n = |b_1|^2 E_1 + |b_2|^2 E_2$$

The coefficients satisfy the condition $|b_1|^2 + |b_2|^2 = 1$. The expectation value of the Hamiltonian in this state as a function of $L$ is calculated by means of $E = \langle \psi|H|\psi \rangle$, which result in:

$$E(L) = W(L)[1 + 3\lambda(L) + (3 + 2\lambda(L))|b_2|^2] \tag{12}$$

Setting the expectation value to be equal to $E_L$ i.e. $n = 2$

$$\frac{\pi^2 \hbar^2}{2mL_3^2}[4 + 5\lambda(L)] = \frac{\pi^2 \hbar^2}{2mL^2}[1 + 3\lambda(L) + (3 + 2\lambda(L))|b_2|^2]$$

$$L^2 = \frac{L_3^2}{[4+5\lambda(L)]}[1 + 3\lambda(L) + (3 + 2\lambda(L))|b_2|^2] \tag{13}$$

The max value of $L$ is when $L = L_4$ and this is achieved in the isothermal expansion when $|b_2|^2 = 0$. Therefore, from equ. (13)

$$L_4 = L_3[1 + 3\lambda(L)/4 + 5\lambda(L)]^{½} \tag{14}$$

The pressure during the isothermal compression is:

$$P_3(L) = \frac{\pi^2 \hbar^2}{mLL_3^2}[4 + 5\lambda] \tag{15}$$

The product $LP_3(L) = constant$. This is an exact quantum analogue of a classical *equation of state*.

**Process 4: Adiabatic Compression**

The system returns to the initial state $n = 1$ at point 4 as it compresses adiabatically (i.e. from $L = L_4$ until $L = L_1$). The expectation of the Hamiltonian is given by:

$$E_H = 2W(L)[1 + 3\lambda(L)] \tag{16}$$

and the pressure applied to the potential well's wall $P$ as a function of $L$ is:

$$P_4(L) = \frac{2W(L)}{L}[1 + 3\lambda(L)\{1 - \mu(L)\}] \equiv P_1(L) \tag{17}$$

The work $W$ performed by the quantum heat engine during one closed cycle, along the four processes described above is the area of the closed loops represented in the *Fig.1*. By using eqs. (9), (11), (15) and (17) one obtains

$$W = W_{12} + W_{23} + W_{34} + W_{41}$$

$$W = \int_{L_1}^{L_2} P_1 dL + \int_{L_2}^{L_3} P_2(L) dL + \int_{L_3}^{L_4} P_3 dL + \int_{L_4}^{L_1} P_4(L) dL$$

Recall that;

$$L_2 = L_1[4 + 5\lambda(L)/1 + 3\lambda(L)]^{½}$$

$$L_4 = L_3[1 + 3\lambda(L)/4 + 5\lambda(L)]^{½}$$

Therefore;

$$W = \int_{L_1}^{L_1[4+5\lambda(L)/1+3\lambda(L)]^{½}} \frac{\pi^2\hbar^2}{mLL_1^2}[1 + 3\lambda]dL + \int_{L_1[4+5\lambda(L)/1+3\lambda(L)]^{½}}^{L_3} \frac{\pi^2\hbar^2}{mL^3}[4 + 5\lambda(L)\{1 - \mu(L)\}]dL$$

$$+ \int_{L_3}^{L_3[1+3\lambda(L)/4+5\lambda(L)]^{½}} \frac{\pi^2\hbar^2}{mLL_3^2}[4 + 5\lambda]dL + \int_{L_3[1+3\lambda(L)/4+5\lambda(L)]^{½}}^{L_1} \frac{\pi^2\hbar^2}{mL^3}[1 + 3\lambda(L)\{1 - \mu(L)\}]dL$$

$$W = \frac{\pi^2\hbar^2}{mL_1^2}\left[\frac{[1+3\lambda]^{3/2}}{[4+5\lambda]^{½}} - [1 + 3\lambda]^{½}[4 + 5\lambda]^{½}\right] - \frac{\pi^2\hbar^2}{mL_3^2}\left[\frac{[4+5\lambda]^{3/2}}{[1+3\lambda]^{½}} - [1 + 3\lambda]^{½}[4 + 5\lambda]^{½}\right]$$

(18)

The efficiency of the heat engine is defined to be:

$$\eta = \frac{W}{Q_H} \tag{19}$$

given that $Q_H$ is the quantity of heat in the hot reservoir and $W$ is the work performed by the classical heat engine. Where $Q_H$ is the heat engine absorbed by the potential well during the isothermal expansion in quantum engine:

$$Q_H = \frac{\pi^2\hbar^2}{mL_1^2}\left[\frac{[1+3\lambda]^{3/2}}{[4+5\lambda]^{½}} - [1 + 3\lambda]\right] \tag{20}$$

Therefore, the efficiency $\eta$ of a quantum heat engine can be defined using (20) as:

$$\eta = 1 - \frac{L_1^2}{L_3^2}\left[\frac{4+5\lambda}{1+3\lambda}\right] \qquad (21)$$

Substituting the Eqs. (12) and (19) into (21), the efficiency can be written as:

$$\eta = 1 - \frac{E_C}{E_H}. \qquad (22)$$

Note that this efficiency is analogous to that of a classical thermodynamic Carnot cycle

## 3. OUR RESULT

In order to validate our result, the derived efficiency in equation (21) needs to be compared with earlier works of *Bender et al* [19] and *Abe* [26] who examined the Carnot engine as a quantum particle in a potential well i.e. free particle [FP] in the box model and Harmonic Oscillator[ HO] model respectively[27].

In the free particle [FP] in the box model, $(\lambda(L) \cong 0)$ at the limit [1,27]. Therefore, the efficiency reduces to:

$$\eta = 1 - \left(2\frac{L_1}{L_3}\right)^2 \qquad (23)$$

while in the Harmonic Oscillator [HO] model, $(\lambda(L) \to 1)$ at the limit, we obtained:

$$\eta = 1 - \left(\frac{3}{2}\frac{L_1}{L_3}\right)^2 \qquad (24)$$

After inserting the necessary conditions, the derived efficiency for the free particle [FP] in the box and Harmonic Oscillator [HO] model is exactly the same as in [19] and [26] respectively which is analogous to classical Carnot cycle.

## CONCLUSION

In this work, we showed that the Pöschl-Teller [PT] oscillator can be used as a working fluid in a quantum engine, we showed that it's possible to construct equations that are analogous to classical adiabatic and isothermal process using the Pöschl-Teller [PT] model. We found that the efficiency obtained in this work agrees with those of earlier studies and it is analogous to the well-known efficiency from classical thermodynamics.

## REFERENCES

[1]   Y.G. Rudoy, E.O. Oladimeji, Pressure Operator for the Pöeschl-Teller Oscillator, Rudn J. Math. Inf. Sci. Phys. 25 (2017) 276–282. https://doi.org/10.22363/2312-9735-2017-25-3-276-282.

[2]   Y.G. Rudoy, E.O. Oladimeji, About one interesting and important model in quantum mechanics I. Dynamic decription, Phys. High. Educ. 23 (2017) 20–32. http://arxiv.org/abs/1906.02274.

[3]   Y.G. Rudoy, E.O. Oladimeji, About One Interesting and Important Model in Quantum Mechanics II. Thermodynamic Description, Phys. High. Educ. 23 (2017) 11–23. http://arxiv.org/abs/1904.09830.

[4]   E.O. Oladimeji, The efficiency of quantum engines using the Pöschl – Teller like oscillator model, Phys. E Low-Dimensional Syst. Nanostructures. 111 (2019) 113–117. https://doi.org/10.1016/j.physe.2019.03.002.

[5]   H.E.D. Scovil, E.O. Schulz-DuBois, Three-Level Masers as Heat Engines, Phys. Rev. Lett. 2

(1959) 262–263. https://doi.org/10.1103/PhysRevLett.2.262.

[6]  R. Kosloff, A quantum mechanical open system as a model of a heat engine, J. Chem. Phys. 80 (1984) 1625–1631. https://doi.org/10.1063/1.446862.

[7]  T. Feldmann, E. Geva, R. Kosloff, P. Salamon, Heat engines in finite time governed by master equations, Am. J. Phys. 64 (1996) 485. https://doi.org/10.1119/1.18197.

[8]  E. Geva, R. Kosloff, A quantum-mechanical heat engine operating in finite time. a model consisting of spin-1/2 systems as the working fluid, J. Chem. Phys. 96 (1992) 3054–3067. https://doi.org/10.1063/1.461951.

[9]  A. Sisman, H. Saygin, On the power cycles working with ideal quantum gases: I. The Ericsson cycle, J. Phys. D. Appl. Phys. 32 (1999) 664–670. https://doi.org/10.1088/0022-3727/32/6/011.

[10]  T. Feldmann, R. Kosloff, Performance of discrete heat engines and heat pumps in finite time, Phys. Rev. E. 61 (2000) 4774–4790. https://doi.org/10.1103/physreve.61.4774.

[11]  T. Feldmann, R. Kosloff, Quantum four-stroke heat engine: Thermodynamic observables in a model with intrinsic friction, Phys. Rev. E - Stat. Physics, Plasmas, Fluids, Relat. Interdiscip. Top. 68 (2003) 18. https://doi.org/10.1103/physreve.68.016101.

[12]  T. Feldmann, R. Kosloff, Characteristics of the limit cycle of a reciprocating quantum heat engine, Phys. Rev. E - Stat. Physics, Plasmas, Fluids, Relat. Interdiscip. Top. 70 (2004) 13. https://doi.org/10.1103/physreve.70.046110.

[13]  R.U. and R. Kosloff, The multilevel four-stroke swap engine and its environment, New J. Phys. 16 (2014) 95003. http://stacks.iop.org/1367-2630/16/i=9/a=095003.

[14]  A.S.L.M. and A.J.S. and P. Kammerlander, A.J. Short, Philipp, Clock-driven quantum thermal engines, New J. Phys. 17 (2015) 45027. http://stacks.iop.org/1367-2630/17/i=4/a=045027.

[15]  M.B. and A.X. and A.F. and G.D.C. and N.K. and M. Paternostro, A. Xuereb, A. Ferraro, G. De Chiara, N. Kiesel, M, Out-of-equilibrium thermodynamics of quantum optomechanical systems, New J. Phys. 17 (2015) 35016. http://stacks.iop.org/1367-2630/17/i=3/a=035016.

[16]  H.T. Quan, Y.X. Liu, C.P. Sun, F. Nori, Quantum thermodynamic cycles and quantum heat engines, Phys. Rev. E. 76 (2007) 31105. https://doi.org/10.1103/PhysRevE.76.031105.

[17]  J. He, X. He, W. Tang, The performance characteristics of an irreversible quantum Otto harmonic refrigeration cycle, Sci. China, Ser. G Physics, Mech. Astron. (2009). https://doi.org/10.1007/s11433-009-0169-z.

[18]  C.L. Chen, C.E. Ho, H.T. Yau, Performance analysis and optimization of a solar powered stirling engine with heat transfer considerations, Energies. 5 (2012) 3573–3585. https://doi.org/10.3390/en5093573.

[19]  C.M. Bender, D.C. Brody, B.K. Meister, Quantum mechanical Carnot engine, J. Phys. A. Math. Gen. 33 (2000) 4427–4436. https://doi.org/10.1088/0305-4470/33/24/302.

[20]  L. Guzmán-Vargas, Efficiency of simple quantum engines: The Joule-Brayton and Otto cycles, Phys. Rev. E - Stat. Nonlinear, Soft Matter Phys. 643 (2003) 291–296. https://doi.org/10.1063/1.1523819.

[21]  M.O. Scully, Extracting Work from a Single Heat Bath via Vanishing Quantum Coherence II: Microscopic Model, in: AIP Publishing, 2003: pp. 83–91. https://doi.org/10.1063/1.1523786.

[22]   H.T. Quan, P. Zhang, C.P. Sun, Quantum-classical transition of photon-Carnot engine induced by quantum decoherence, Phys. Rev. E. 73 (2006) 036122. https://doi.org/10.1103/PhysRevE.73.036122.

[23]   M.J. Henrich, G. Mahler, M. Michel, Driven spin systems as quantum thermodynamic machines: Fundamental limits, Phys. Rev. E. 75 (2007) 051118. https://doi.org/10.1103/PhysRevE.75.051118.

[24]   R. Tah, Simulating a Quantum Harmonic Oscillator by introducing it to a Bosonic System, 2020. https://doi.org/10.35543/osf.io/xh8wf.

[25]   R. Kosloff, Y. Rezek, The Quantum Harmonic Otto Cycle, Entropy. 19 (2017) 1–36. https://doi.org/10.3390/e19040136.

[26]   S. Abe, General Formula for the Efficiency of Quantum-Mechanical Analog of the Carnot Engine, Entropy. 15 (2013) 1408–1415. https://doi.org/10.3390/e15041408.

[27]   S.-H.H. Dong, Factorization Method in Quantum Mechanics, 1st ed., Springer, Dordrecht, 2007. https://doi.org/10.1007/978-1-4020-5796-0.

[28]   R. Kosloff, A. Levy, Quantum heat engines and refrigerators: Continuous devices, Annu. Rev. Phys. Chem. 65 (2013) 365–393. https://doi.org/10.1146/annurev-physchem-040513-103724.

[29]   M.D. Lechner, Einführung in die Quantenchemie. Von H. Hellmann. 350 S.,43 Abb., 35 Tab. Franz Deuticke, Leipzig u. Wien 1937. Pr. geh. RM. 20,-. geb. RM. 22,-, Franz Deuticke, 2007. https://doi.org/10.1002/ange.19410541109.

[30]   R.P. Feynman, Forces in molecules, Phys. Rev. 56 (1939) 340–343. https://doi.org/10.1103/PhysRev.56.340.

[31]   G. Pöschl, E. Teller, Bemerkungen zur Quantenmechanik des anharmonischen Oszillators, Zeitschrift für Phys. 83 (1933) 143–151. https://doi.org/10.1007/BF01331132.

[32]   S. Raghavan,  a. Bishop, V. Kenkre, Quantum versus semiclassical description of self-trapping: Anharmonic effects, Phys. Rev. B. 59 (1999) 9929–9932. https://doi.org/10.1103/PhysRevB.59.9929.

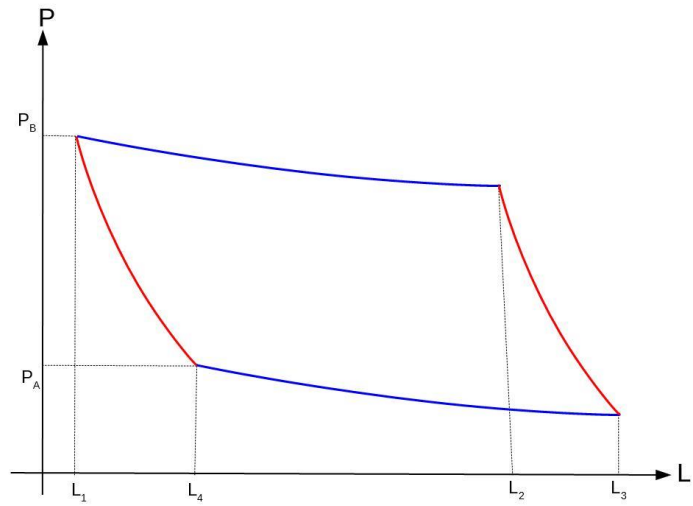

*Figure 1: The schematic representation of the Carnot's cycle.1*